\documentclass[aps,11pt,nofootinbib,showpacs,preprintnumbers]{revtex4-2}

\usepackage{float}
\usepackage{graphicx}
\usepackage{subcaption}
\usepackage{amsmath, amssymb, graphicx, hyperref}
\usepackage{xcolor}
\usepackage[margin=1in]{geometry}

\newcommand{\be}{\begin{equation}} 
\newcommand{\ee}{\end{equation}} 
\newcommand{\ba}{\begin{eqnarray}} 
\newcommand{\ea}{\end{eqnarray}}

\newcommand{\bea}{\begin{eqnarray}} 
\newcommand{\eea}{\end{eqnarray}}

\begin{document}

\title{Cosmological Dynamics of the Thermal Scalar Near the Hagedorn Temperature}

\author{Arnab Pradhan$^{1,a}$, Luis Rufino$^{2,b}$, and Scott Watson$^{2,c}$}

\affiliation{$^{1}$Department of Physics and Astronomy, York University, Toronto, Ontario M3J 1P3, Canada}

\affiliation{$^{2}$Department of Physics, Syracuse University, Syracuse, New York 13244, USA}

\email[$^{a}$]{arpradha@yorku.ca}
\email[$^{b}$]{lerufino@syr.edu}
\email[$^{c}$]{gswatson@syr.edu}
\date{\today}

\begin{abstract}
We study the cosmological dynamics of the thermal scalar—the winding-string mode that becomes massless at the Hagedorn transition—by coupling it to the string-frame gravi-dilaton effective action. This provides a field-theoretic framework for investigating winding-mode dynamics near the Hagedorn temperature and their role in string cosmology.

Below the Hagedorn temperature, the phase space contains static configurations in which the thermal scalar balances the shifted dilaton evolution. These configurations are boundary states rather than attractors, and when the thermal-scalar mass depends on the scale factor, winding-mode back-reaction opposes expansion and can reverse it. Above the Hagedorn temperature, the tachyonic thermal scalar generates negative effective energy density while preserving the null energy condition, enabling branch changes of the Brustein–Veneziano type. These transitions do not provide the graceful exit required to connect the Hagedorn phase to standard cosmological evolution. At the Hagedorn temperature itself, the quadratic effective theory breaks down and higher-order interactions become essential.

Because the thermal scalar originates as a Euclidean order parameter, our Lorentzian treatment should be viewed as an effective dynamical model of the Hagedorn transition. Within this framework, the thermal scalar clarifies the dynamical structure surrounding the Hagedorn transition and shows how the Hagedorn exit problem cannot be resolved within the quadratic effective theory alone.

\end{abstract}

\maketitle
\newpage 

\section{Introduction}
String Gas Cosmology was proposed as a framework for addressing the initial cosmological singularity and fine-tuning problems of inflationary cosmology~\cite{Brandenberger:1988aj,Battefeld:2005av}. The picture, originally proposed by Brandenberger and Vafa~\cite{Brandenberger:1988aj}, begins with a small, hot, compact space, filled with a gas of fundamental strings in thermal equilibrium -- seeking an alternative to inflation. As the universe is compressed toward the string scale, the exponential growth of the density of states causes the temperature to saturate at a finite limiting value, the Hagedorn temperature $T_H$~\cite{Hagedorn:1965st}. The universe enters a quasi-static phase --- a lingering state~\cite{Melcher:2023kpd} --- at finite size and finite temperature. Because winding and anti-winding strings can generically find each other and annihilate only in three or fewer spatial dimensions, this phase provides a mechanism for three dimensions to grow large while the rest remain compact~\cite{Brandenberger:1988aj, Brandenberger_2002}.

The dynamics of a string gas in a cosmological background were studied by Kaloper and Watson~\cite{Kaloper:2007pw}, who showed that the sign of the shifted dilaton velocity $\dot\varphi = \dot\phi - d\dot\lambda$ is conserved whenever the energy density is non-negative. This divides cosmological solutions into two disconnected branches (super-selection sectors): the $(+)$ branch ($\dot\varphi > 0$), on which the string coupling grows and the geometry evolves toward a future inflationary pole singularity, and the $(-)$ branch ($\dot\varphi < 0$), which contains standard radiation-dominated FLRW cosmology. The Hagedorn phase naturally lives on the $(+)$ branch, while the late universe we observe lives on the $(-)$ branch; the singularity separating them is what the authors termed a \textit{geometric precipice}~\cite{Kaloper:2007pw}. Crossing from $(+)$ to $(-)$ --- the graceful exit~\cite{Brustein_1994} --- requires violation of the null energy condition and remains an open problem.

The thermodynamic treatment of~\cite{Kaloper:2007pw} describes the string gas through equations of state but does not resolve the dynamics of individual winding or momentum modes. A field-theoretic description of the relevant light mode was developed by Atick and Witten~\cite{Atick:1988si}, who showed that the Hagedorn transition is associated with a winding mode on the thermal circle becoming tachyonic, and further by Horowitz and Polchinski~\cite{Horowitz_1998}, who identified the thermal scalar $\chi$ --- the lightest winding mode --- with a temperature-dependent mass $\mu^2(\beta) \propto \beta^2 - \beta_H^2$. Below $T_H$ the thermal scalar is massive ($\mu^2 > 0$), at $T_H$ it is massless ($\mu^2 = 0$), and above $T_H$ it becomes tachyonic ($\mu^2 < 0$). In this work, we couple the thermal scalar to the low-energy gravi-dilaton effective action and study the resulting cosmological dynamics. In addition, one can derive an analogous field $\psi$ describing the lowest momentum modes in a similar fashion. The interaction between winding and momentum sectors is expected to play a role in the transition out of the Hagedorn phase; we return to this in Section~\ref{sec:momentummodes}.

A further motivation for developing a field-theoretic description of the Hagedorn phase comes from the work of Dienes,  Lennek and Sharma~\cite{Dienes_2012}, who consider the role of Wilson lines on the Hagedorn temperature for different gauge sectors and theory types (A/B). Understanding the dynamics of the Hagedorn transition at the level of field theory without Wilson lines is a necessary first step toward studying the cosmological implications of this additional structure, which we plan to address in future work.

The classical dynamics of the lingering-to-inflation transition were studied in~\cite{Melcher:2023kpd,Melcher:2025vku} within general relativity, where it was shown that a universe beginning in a stalled phase can exit to inflation entirely on the $(+)$ branch, avoiding the need for a branch change and thus preserving the null energy condition throughout. In particular, in this analysis it was shown that the flow from arbitrary initial conditions leads toward the fixed point of a lingering universe, but it is a hyperbolic fixed point. That is, the Hagedorn transition occurs as a hyperbolic fixed point and therefore is not stable -- although the field can `linger'. 

The present work addresses a complementary question from the string-theoretic side: what does the field-theoretic description of the thermal scalar tell us about the dynamics \emph{within} the Hagedorn regime itself? We map out the phase space structure across the three mass regimes, identify the configurations that realize lingering  and show that the tachyonic regime realizes, within the thermal-scalar effective theory, the conditions for branch changing. The exit from the Hagedorn phase --- which requires winding-number violation and thus physics beyond the quadratic truncation --- is left for future work incorporating the quartic action of~\cite{Brustein_2021}.

This paper is organized as follows. In Section~\ref{review} we review the gravi-dilaton cosmological framework and the branch structure of string cosmology.   In Section~\ref{sec:thermalscalar} we review the branch structure of string cosmology following~\cite{Kaloper:2007pw, Brustein_1994} and couple the thermal scalar to the gravi-dilaton action. In Section~\ref{sec:phasespace} we perform the phase-space analysis across the three mass regimes. In Section~\ref{sec:momentummodes} we discuss the momentum-mode sector and explain why it is not included in the present analysis. We conclude in Section~\ref{sec:discussion}.

\section{Hagedorn Cosmology and branch structure}\label{review}
This section reviews the gravi-dilaton cosmological framework of~\cite{Kaloper:2007pw, Brustein_1994} that underlies our analysis; readers familiar with this material may proceed directly to Section~\ref{sec:thermalscalar}. The key results are the branch equation~\eqref{eq:branchequation} and the obstruction to branch changes encoded in~\eqref{eq:integratedbranchchange}.

The authors of~\cite{Kaloper:2007pw} show that the Hamiltonian constraint of string cosmology can be rearranged into the form
\begin{equation}
    \dot\varphi_s = \pm\sqrt{d H_s^2 + e^{\phi_s}\rho_s}\,,
    \qquad \varphi_s = \phi_s - d\lambda_s\,,
    \label{eq:branchequation}
\end{equation}
where $\lambda_s = \ln a$ is the logarithmic scale factor, $d$ is the number of space dimensions, $\phi_s$ is the dilaton, and $H_s = \dot\lambda_s$ is the string-frame Hubble rate. For non-negative energy density the argument of the square root is strictly positive, so the sign of $\dot\varphi_s$ cannot change continuously: it is fixed by the initial conditions and divides the space of solutions into two disconnected branches.

The mechanism by which branch changes can occur was first identified by Brustein and Veneziano \cite{Brustein_1994}, who showed that a dilaton potential $V(\phi)$ with a negative region creates a boundary in phase space -- the "Egg", defined by $3H^2 + U(\phi) = 0$ with $U = e^{\phi}V$, where trajectories oscillate and switch branches. However they also show that such branch changes generically come in pairs. A trajectory that touches the Egg and switches from $(+) \to (-)$ will unavoidably hit the Egg again and return to the $(+)$ branch, preventing a complete graceful exit in the weak-curvature regime. 

Kaloper and Watson~\cite{Kaloper:2007pw} showed that when $\rho_s \geq 0$, no branch change is possible. If $\rho_s < 0$, branch changes from $(-) \to (+)$ can occur without violating NEC. However, connecting the Hagedorn phase to a standard expanding FLRW cosmology requires the reverse transition $(+) \to (-)$, which demands NEC violation~\cite{Kaloper:2007pw, Melcher:2023kpd}.

\subsection{Dynamics and Branch Structure}
\label{subsec:keyeqs}

At energies well below the string scale, string theory reduces to an effective field theory. The universal massless sector contains the spacetime metric $g_{\mu\nu}$ and a scalar field $\phi_s$ called the dilaton, whose expectation value sets the string coupling $g_s = e^{\phi_s}$. The low-energy dynamics of these fields in the string frame is governed by the $(d+1)$-dimensional effective action
\begin{equation}
    S = \frac{1}{2\kappa_d}\int d^{d+1}x\,\sqrt{-g_s}\,
    e^{-\phi_s}\Big(R_s + (\partial\phi_s)^2 
    - \mathcal{L}_m\Big)\,,
    \label{eq:stringaction}
\end{equation}
where $\mathcal{L}_m$ is the matter Lagrangian and we have set the Kalb--Ramond two-form to zero throughout. For homogeneous, possibly anisotropic backgrounds, the metric takes the form
\begin{equation}
    ds^2 = -n(t)^2\,dt^2 + \sum_{i=1}^d 
    e^{2\lambda_s^{(i)}(t)}\,dx_i^2\,,
    \label{eq:metricansatz}
\end{equation}
with $n(t)$ the lapse function, and the shifted dilaton is defined as
\begin{equation}
    \varphi_s = \phi_s - \sum_{i=1}^d \lambda_s^{(i)}\,,
    \label{eq:shifteddilaton}
\end{equation}
whose time derivative encodes the branch structure through Eq.~\eqref{eq:branchequation}. Varying the action yields the Hamiltonian constraint, scale factor equations, dilaton equation, and conservation law; in the isotropic limit $\lambda_s^{(i)} \equiv \lambda_s$ these reduce to a string-frame Friedmann system from which the branch equation~\eqref{eq:branchequation} follows directly. The full equations of motion are derived in~\cite{Kaloper:2007pw}.

The matter sector is parameterized by the equation-of-state variable
\begin{equation}
    \gamma = \frac{p_s}{\rho_s}\,,
    \label{eq:eosgamma}
\end{equation}
with the principal string-gas regimes given by $\gamma = 0$ (Hagedorn phase), $\gamma = 1/d$ (momentum-mode domination), and $\gamma = -1/d$ (winding-mode domination).

The obstruction to branch changing can be made precise by integrating the equations of motion between an initial Hagedorn configuration at $t_h$ and a final expanding configuration at $t_e$~\cite{Kaloper:2007pw}:
\begin{equation}
    \sqrt{d-1}\,H_s(t_e) + H_s(t_h) + A = 
    -\frac{1}{2}\int_{t_h}^{t_e} dt\,
    e^{\phi_s}(\rho_s + p_s)\,,
    \label{eq:integratedbranchchange}
\end{equation}
from which a branch change requires
\begin{equation}
    \rho_s + p_s < 0\,.
    \label{eq:necviolation}
\end{equation}
Branch changes are therefore forbidden for matter obeying the null energy condition.

However, the transition from a lingering configuration to inflation remains on the $(+)$ branch and does not encounter the geometric precipice. The physical mechanism that could mediate such a transition --- the annihilation of winding modes into string loops~\cite{Brandenberger_2002} --- lies beyond the quadratic truncation studied here, as discussed in Section~\ref{sec:momentummodes}.

\section{Thermal Scalar}
\label{sec:thermalscalar}
The thermal scalar $\chi$ is a complex scalar describing the $\pm 1$ winding modes of closed strings on the Euclidean thermal circle $S_{\beta}^1$. Atick and Witten~\cite{Atick:1988si} showed that the Hagedorn transition is associated with this winding mode becoming tachyonic, and the effective field theory description we use throughout this paper is due to Horowitz and Polchinski~\cite{Horowitz_1998}, who identified a complex scalar of unit winding number with a temperature-dependent mass. 
\begin{equation}
\mu^2(\beta) = \frac{\beta^2 - \beta_H^2}{(2\pi\alpha')^2}\,,
\label{eq:thermalmass}
\end{equation}
where $\beta = T^{-1}$, $\beta_H = T_H^{-1}$, and $\alpha'$ is related to the fundamental string length as $l_s = \sqrt{\alpha^\prime}$.  The mass is positive below $T_H$, vanishes at $T_H$, and becomes tachyonic above. The vanishing of $\mu^2$ at the critical temperature makes $\chi$ the light degree of freedom controlling the transition, which is why we take its dynamics as the microscopic input in what follows.

The thermal scalar is not a conventional propagating field: it arises from the one-loop closed-string partition function as the mode whose vanishing mass at $\beta_H$ signals the Hagedorn divergence~\cite{BARB_N_2005}. Horowitz and Polchinski~\cite{Horowitz_1998} noted that the thermal scalar has no direct dynamical role but serves to characterize the static properties of highly excited strings. Using it as a dynamical field evolving in a cosmological background, as we do here, is therefore itself a nontrivial step --- one whose justification rests on the effective action reproducing the correct critical thermodynamics, not on $\chi$ being a particle in the usual sense.

\subsection{Action}
\label{subsec:thermalaction}
The thermal scalar coupled to the gravi-dilaton sector is described by (all quantities are in the string frame unless otherwise noted)
\begin{equation}
    S = \frac{1}{2\kappa_d^2}\int d^{d+1}x\,\sqrt{-g}\, 
    e^{-\phi}\left(-R - (\partial\phi)^2 + |\partial\chi|^2 
    - \mu^2(\beta)\,|\chi|^2\right),
    \label{eq:thermalaction}
\end{equation}
with $\kappa_d^2 = 8\pi G_d$. The winding sector $\mathcal{L}_\chi$ shares the same $e^{-\phi}$ prefactor as the gravi-dilaton sector: if $\chi$ is the order parameter for the Hagedorn transition in a cosmological setting, the dilaton must couple to the winding sector and its mass. This is the form studied in~\cite{Brustein_2021}. Decoupling $\chi$ from the gravi-dilaton sector would remove the dilaton friction term $-\dot\varphi\,\dot\chi$ from the $\chi$ equation of motion, giving qualitatively different dynamics.

We work with the homogeneous, isotropic metric
\begin{equation}
    ds^2 = n(t)^2\,dt^2 - e^{2\lambda(t)}\,
    d\vec{x}^{\,2}_{d}\,,
    \label{eq:metricansatz2}
\end{equation}
where $\lambda(t) = \ln a(t)$, $n(t)$ is the lapse function, and we assume the extra dimensions are compactified and frozen, working in the resulting $3+1$-dimensional effective theory ($d = 3$). The dilaton $\phi$ and the thermal scalar $\chi$ are taken to be spatially homogeneous.

\subsection{Equations of motion}
\label{subsec:thermaleom}
Inserting the homogeneous-isotropic metric~\eqref{eq:metricansatz2} into the action~\eqref{eq:thermalaction}, we obtain
\begin{equation}
S_{1D} = \frac{V_d}{2\kappa_d^2} \int dt\, n\, e^{-\varphi}
\left[\frac{1}{n^2}\left(-\dot\varphi^2 +d\dot\lambda^2 +
\dot\chi^2\right) - \mu^2\chi^2\right]\,,
\label{eq:1Daction}
\end{equation}
where $V_d$ is the coordinate volume of the $d$-dimensional spatial slice, and $\varphi = \phi - d\lambda$. 
The shifted dilaton $\varphi$ is a convenient combination that absorbs the scale factor evolution into the dilaton prefactor, simplifying the resulting equations.

Varying~\eqref{eq:1Daction} with respect to the lapse and then fixing the gauge to $n = 1$ gives the Hamiltonian constraint
\begin{equation}
    \dot\varphi^2 - d\dot\lambda^2 = \dot\chi^2 + \mu^2\chi^2\,.
    \label{eq:thermalconstraint}
\end{equation}
Varying with respect to $\lambda$, $\varphi$, and $\chi$ yields the scale factor equation
\begin{equation}
    \ddot\lambda - \dot\varphi\,\dot\lambda =
    -\frac{1}{d}\frac{d\mu^2}{d\lambda}\,\chi^2\,,
    \label{eq:thermallambda}
\end{equation}
the shifted dilaton equation (after using the Hamiltonian constraint)
\begin{equation}
    \ddot\varphi = \dot\varphi^2 - \mu^2\chi^2\,,
    \label{eq:thermalvarphireduced}
\end{equation}
and the thermal scalar equation of motion
\begin{equation}
    \ddot\chi - \dot\varphi\,\dot\chi + \mu^2\chi = 0\,.
    \label{eq:thermalchi}
\end{equation}
Several features of this system are worth noting. First, the $\dot\varphi\,\dot\chi$ term in Eq.~\eqref{eq:thermalchi} shows that the dilaton acts as a source of friction or anti-friction on the thermal scalar, depending on the sign of $\dot\varphi$. On the $(-)$ branch ($\dot\varphi < 0$), this term provides positive friction that damps the oscillations of $\chi$; on the $(+)$ branch ($\dot\varphi > 0$), it acts as anti-friction. Second, the scale factor is sourced by $\chi$ only when $\mu^2$ depends on $\lambda$; if $\mu$ was a constant, the right-hand side of Eq.~\eqref{eq:thermallambda} would vanish and the scale factor would decouple from the thermal scalar at the level of the scale factor evolution. Third, the absence of $e^{\varphi}$ factors on the right-hand sides of Eqs.~\eqref{eq:thermalconstraint}--\eqref{eq:thermalchi} distinguishes this system from the corresponding equations in~\cite{Kaloper:2007pw}, where the matter free energy does not share the dilaton prefactor. The consequence of this difference becomes visible in the static limit: our static solutions carry an equation of state $\gamma = -1$, whereas~\cite{Kaloper:2007pw} assigns $\gamma = 0$ to the corresponding Hagedorn phase. We return to this in Section~\ref{sec:massive}.

The coupled system does not admit closed-form solutions in general. To develop intuition for the dynamics, it is useful to begin with the Hamiltonian constraint~\eqref{eq:thermalconstraint}. Defining the string-frame Hubble parameter $H_s = \dot\lambda$, the constraint can be written as
\begin{equation}
    H_s = \pm\sqrt{\frac{\dot{\varphi}^2 - \rho_\chi}{d}}\,,
    \qquad \rho_{\chi} = \dot{\chi}^2 + \mu^2(\beta)\chi^2\,.
    \label{eq:branchconstraint}
\end{equation}

It is important to distinguish the effective energy density $\rho_\chi$ from the matter energy density appearing in the branch-change analysis of \cite{Kaloper:2007pw} because the thermal scalar carries the same dilaton prefactor as the gravi-dilaton sector, its contribution to the Hamiltonian constraint differs from that of the free-energy source considered in String Gas Cosmology. Consequently, the analogy with branch changing should be understood as a statement about the phase-space structure of the effective theory rather than as a direct application of the theorem of \cite{Kaloper:2007pw}.

This equation is central to the analysis because it determines the region of phase space accessible to the cosmological evolution and makes the branch structure of the shifted dilaton manifest. For $H_s$ to be real, the shifted dilaton velocity must satisfy $\dot\varphi^2 \geq \rho_\chi$; configurations that violate this bound lie outside the physical phase space. The boundary $\dot\varphi^2 = \rho_\chi$ corresponds to $H_s = 0$ --- the static solutions\footnote{The terminology in the string cosmology literature is not uniform: configurations with $H_s = 0$ are variously called ``loitering''~\cite{Kaloper:2007pw} or ``lingering''~\cite{Melcher:2023kpd}, sometimes with overlapping and sometimes with distinct meanings. We use \emph{static} for any configuration with $H_s = 0$. We reserve \emph{loitering} for the degenerate case $\mu^2 = 0$ in which the fixed points collapse to the origin $(\dot\varphi, H_s) = (0,0)$, and \emph{lingering} for the static fixed points of the $\mu^2 > 0$ regime at $(\dot\varphi, H_s) = (\pm\mu|\chi|, 0)$ with $\dot\chi = 0$.}.

\begin{figure}[htbp]
    \centering
    \includegraphics[width=0.5\linewidth]{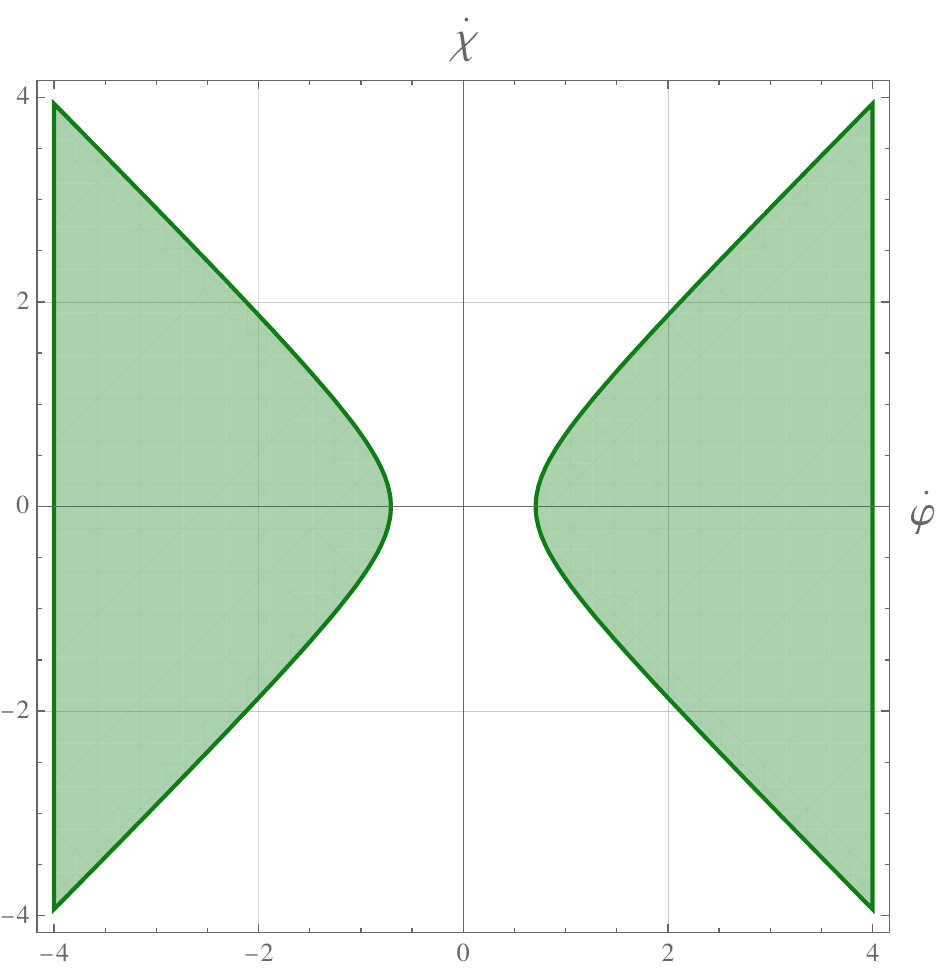}
    \caption{Allowed region of phase space in the $(\dot\varphi, \dot\chi)$. The boundary of the shaded region corresponds to $H_s = 0$, i.e.\ the static configurations. Configurations inside the boundary are physically allowed; those outside have $\dot\varphi^2 < \rho_\chi$ and do not admit real solutions for $H_s$.}
    \label{fig:allowedregions}
\end{figure}

\subsection{Interpretation of the Thermal Scalar}

The thermal scalar occupies a somewhat unusual position in the present analysis. Unlike an ordinary propagating degree of freedom, it originates from the Euclidean finite-temperature partition function and serves as an order parameter for the Hagedorn transition \cite{Atick:1988si, Horowitz_1998}. Its vanishing mass at $T_H$ signals the onset of the Hagedorn instability, and its effective action reproduces the critical thermodynamics of highly excited strings. However, the thermal scalar is not generally interpreted as a physical Lorentzian particle whose real-time evolution can be derived directly from string theory.

Throughout this work we adopt a more modest viewpoint. We treat the thermal scalar as an effective dynamical proxy for the winding sector and study the cosmological phase-space structure that results from coupling this field to the gravi-dilaton system. The resulting equations should therefore be understood as an effective description motivated by string thermodynamics rather than as a controlled Lorentzian derivation from first principles. Our goal is not to establish the real-time dynamics of the thermal ensemble, but rather to investigate what cosmological structures arise if the thermal scalar is used as the light degree of freedom controlling the approach to the Hagedorn transition.

This distinction is particularly important in the tachyonic regime $\mu^2<0$. In that regime the effective equations admit branch-changing trajectories and negative contributions to the Hamiltonian constraint. These features provide a useful dynamical realization of the phase-space structure associated with the Brustein--Veneziano mechanism \cite{Brustein_1994}. However, because the thermal scalar is fundamentally a Euclidean order parameter, such results should be interpreted as properties of the effective theory rather than as controlled predictions of string cosmology.

A second limitation concerns the range of validity of the quadratic truncation. As $\mu^2 \rightarrow 0$, the thermal scalar becomes massless and higher-order interactions become increasingly important. In particular, the quartic couplings identified in Ref.~\cite{Brustein_2021} become comparable to the quadratic terms and eventually dominate the dynamics at the transition itself. The analysis presented below should therefore be viewed as describing the approach to the Hagedorn transition from either side, rather than the transition point itself.

\section{Phase-Space Analysis}
\label{sec:phasespace}
With the equations of motion in hand, we now analyze the cosmological dynamics permitted by the thermal scalar across the three mass regimes. The Hamiltonian constraint $\dot\varphi = \pm\sqrt{d H_s^2 + \rho_\chi}$ plays the role of the Friedmann equation in this string-frame formulation and organizes the possible cosmologies into four classes, distinguished by the signs of $H_s$ and $\dot\varphi$ (Table~\ref{tab:classes}).

On the $(+)$ branch, the string coupling $g_s = e^{\phi}$ grows monotonically. An expanding $(+)$ branch cosmology (Class~I) corresponds to dilaton-driven inflation in the string frame. A contracting $(+)$ branch cosmology (Class~II) is driven toward a future singularity with growing coupling. On the $(-)$ branch, the coupling decreases: an expanding $(-)$ branch cosmology (Class~III) is the standard radiation-dominated FLRW evolution driven by string momentum modes\footnote{As reviewed in \cite{Battefeld:2005av} when winding and anti-winding modes annihilate they produce closed string loops, which as the universe expands, behave as radiation (``momentum modes'').} , while a contracting $(-)$ branch (Class~IV) corresponds to winding-dominated contraction. The dynamics within and between these classes depends on the sign of $\mu^2$; we analyze the three regimes in turn.

\begin{table}[htbp]
    \centering
    \begin{tabular}{c|cc}
        & \textbf{$H_s > 0$ (expanding)} & \textbf{$H_s < 0$ (contracting)} \\
        \hline
        \textbf{$(+)$ branch ($\dot\varphi > 0$)}
        & Class I: dilaton-driven inflation
        & Class II: contracting, growing $g_s$ \\
        \textbf{$(-)$ branch ($\dot\varphi < 0$)}
        & Class III: FLRW expansion
        & Class IV: winding-dominated contraction \\
    \end{tabular}
    \caption{The four classes of cosmology in the $(\dot\varphi, H_s)$ plane, 
    organized by the branch sign of the shifted dilaton and the 
    string-frame Hubble rate. Classes I and II reside on the $(+)$ 
    branch and have future singularities; Classes III and IV reside 
    on the $(-)$ branch and have past singularities.}
    \label{tab:classes}
\end{table}

\subsection{Below the Hagedorn temperature ($\mu^2 > 0$)}
\label{sec:massive}
\begin{figure}[htbp]
    \centering
    \includegraphics[width=0.7\linewidth]{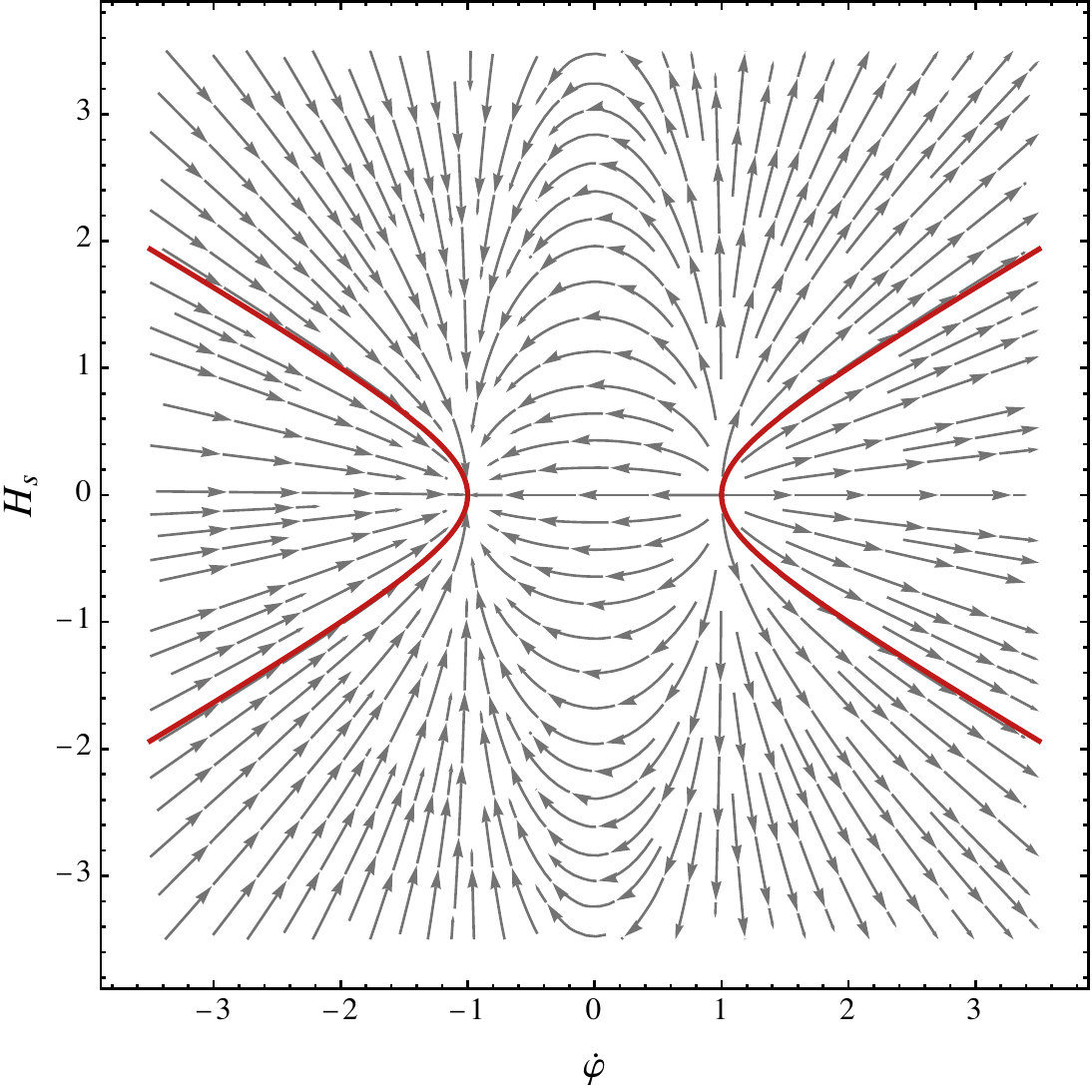}
    \caption{Phase portrait in the $(\dot\varphi, H_s)$ plane
    for $\mu^2 > 0$, $|\chi| = |\dot\chi| = 1$. Fixed points
    sit at $(\pm\mu|\chi|, 0)$; static solutions lie on the
    constraint lines at $(\pm\sqrt{\mu^2\chi^2 + \dot\chi^2}, 0)$.}
    \label{fig:phase-portrait}
\end{figure}

Equation~\eqref{eq:thermalvarphireduced} has fixed points at $(\dot\varphi, H_s) = (\pm\mu|\chi|, 0)$, while static solutions ($H_s = 0$ on the constraint) sit at $(\pm\sqrt{\mu^2\chi^2 + \dot\chi^2}, 0)$. The static solution coincides with a fixed point only when $\dot\chi = 0$; for any $\dot\chi \neq 0$, a trajectory reaching $H_s = 0$ is not a fixed point.

At the static configuration ($\dot\chi = 0$), the thermal scalar sources an energy density $\rho_\chi = \mu^2\chi^2$ and pressure $P_\chi = -\mu^2\chi^2$, giving $\gamma = -1$. This equation of state characterizes the thermal scalar condensate itself; the effective spatial stress driving the scale factor, given by the right-hand side of Eq.~\eqref{eq:thermallambda}, is a distinct quantity discussed below.

The source driving the scale factor evolution is not $P_\chi$ but $(d\mu^2/d\lambda)\chi^2$, which plays the role of an effective spatial pressure and vanishes when $\mu^2$ is $\lambda$-independent. In this case, the scale factor equation reduces to $\ddot\lambda = \dot\varphi\dot\lambda$, which at $H_s = 0$ gives $\ddot\lambda = 0$: the universe is momentarily stalled with vanishing spatial stress, consistent with the $\gamma = 0$ Hagedorn equation of state of~\cite{Kaloper:2007pw} and the near-$\beta_H$ stress tensor analysis of Mertens~\cite{mertens2016hagedornstringthermodynamicscurved}. However, since $\mu^2(\beta) \propto \beta^2 - \beta_H^2$ and $\beta$ can depend on the scale factor, $d\mu^2/d\lambda \neq 0$ generically. 

\begin{figure}[htbp]
    \centering
    \begin{subfigure}[b]{0.48\textwidth}
        \centering
        \includegraphics[width=\textwidth]{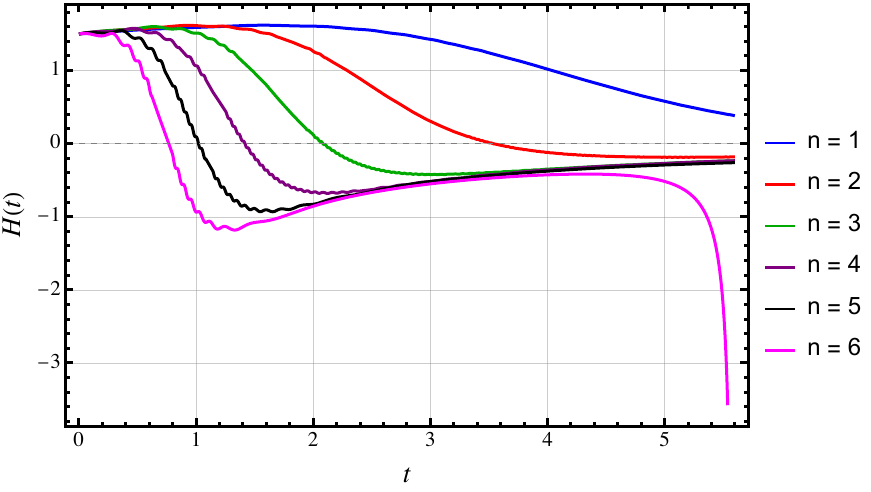}
        \caption{Hubble rate $H_s(t)$ for $n = 1-6$.}
        \label{fig:winding-H}
    \end{subfigure}
    \hfill
    \begin{subfigure}[b]{0.48\textwidth}
        \centering
        \includegraphics[width=\textwidth]{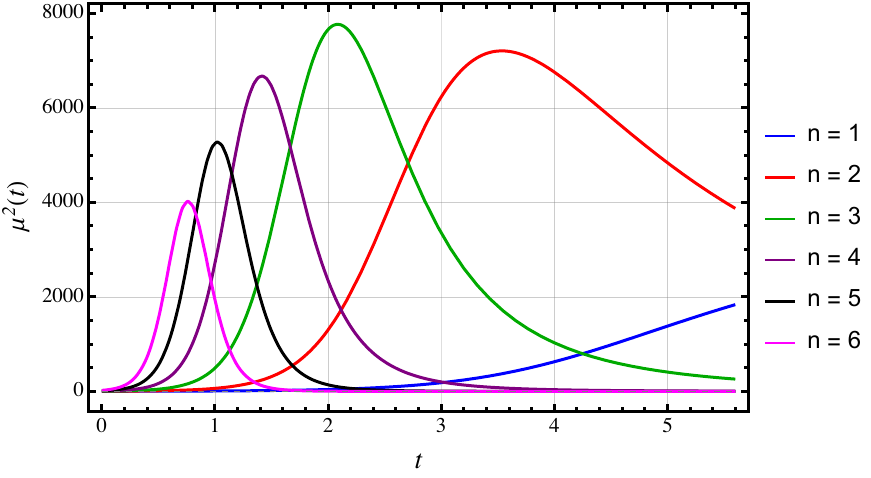}
        \caption{Dynamical mass $\mu^2(t)$ for $n = 1-6$.}
        \label{fig:winding-mu2}
    \end{subfigure}
    \caption{Numerical solutions on the $(+)$ branch for the ansatz $\mu^2(\lambda) = \frac{\beta_H^2}{2}[(a/a_0)^n - 1]$ with initial condition $\lambda_0 = \lambda_s + 0.51$, $\dot\lambda_0 > 0$, $\chi_0 = 0.1$, $\dot\chi_0 = 0$. Left: the winding-mode source reverses the expansion for all $p$, with steeper $\mu^2(\lambda)$ producing faster reversal. Right: as the universe contracts, $\mu^2$ decreases and eventually becomes tachyonic, signaling the onset of the Hagedorn transition.}
    \label{fig:winding-dynamics}
\end{figure}

We illustrate this by solving the coupled system numerically on the $(+)$ branch for the ansatz $\mu^2(\lambda) = \frac{\beta_H^2}{2}\!\left[\left(\frac{a}{a_0}\right)^n - 1\right]$, with $n = 1-6$ (Fig.~\ref{fig:winding-dynamics}). Starting from expanding configurations ($H_s > 0$), the source term $-(1/d)(d\mu^2/d\lambda)\,\chi^2$ in the scale factor equation opposes expansion, reproducing from the thermal scalar field equations the winding-mode back-reaction found on thermodynamic grounds in~\cite{Brandenberger:1988aj, Brandenberger_2002}. For all $n$ the expansion is reversed and the universe enters a contracting phase; steeper $\mu^2(\lambda)$ produces faster reversal.

These results are consistent with the thermodynamic picture of winding-mode back-reaction developed in Refs.~\cite{Brandenberger:1988aj,Brandenberger_2002}. In the thermal-scalar effective theory, the source term $-(1/d)(d\mu^2/d\lambda)\chi^2$ opposes expansion and generically drives initially expanding solutions toward contraction. The numerical solutions further illustrate how rapidly the cosmology is pushed toward a regime of increasing curvature. On the $(+)$ branch, where $H_s>0$ and $\dot{\phi}>0$, both the string coupling and the curvature grow monotonically, driving the system toward the strong-curvature regime $\ell_s^2R\gtrsim1$. This behavior is consistent with the approach to the future singularity identified by Kaloper and Watson~\cite{Kaloper:2007pw} for Class~I winding-mode cosmologies. Once the curvature reaches the string scale, however, the low-energy effective description can no longer be trusted and higher-order $\alpha'$ corrections are expected to become important. The numerical solutions therefore highlight a limitation as much as a result: the quadratic thermal-scalar effective theory captures the approach to the strong-curvature regime but cannot determine the ultimate fate of these trajectories.

The outcome is sensitive to initial conditions. If the universe were to begin at the stalled fixed point with $H_s = 0$ and $\dot\lambda = 0$, and if winding-number-violating interactions --- the annihilation of winding strings with anti-winding strings into momentum modes~\cite{Brandenberger_2002} --- were operative, a transition to a genuinely expanding phase could in principle be initiated. Such interactions are absent in the quadratic action, whose $U(1)$ winding symmetry is an artifact of the truncation. The full string theory is expected to contain no exact global symmetries, and the winding-violating processes that break it are precisely the loop-production interactions modeled in~\cite{Brandenberger_2002}.

As $\mu^2$ approaches zero from above, the two fixed points at $\dot\varphi = \pm\mu|\chi|$ collapse to the origin, and the static configuration that supports the stalled phase becomes increasingly marginal.
At $\mu^2 = 0$ the quadratic action breaks down entirely and the quartic self-interaction~\cite{Brustein_2021} becomes the leading stabilizing contribution.
The thermal scalar description therefore cannot sustain the pressureless Hagedorn phase within the quadratic truncation: the system approaches it but requires additional physics to remain there.

\subsection{Above the Hagedorn temperature ($\mu^2 < 0$)}
\label{sec:tachyonic}

Above $T_H$, the thermal scalar mass $\mu^2 = (\beta^2 - \beta_H^2)/(2\pi\alpha')^2$ becomes negative. This is the field-theoretic signature of the Hagedorn phase transition~\cite{Atick:1988si, BARB_N_2005}: the winding mode becomes massless at $T_H$ and tachyonic above it.

\begin{figure}[htbp]
    \centering
    \includegraphics[width=0.7\linewidth]{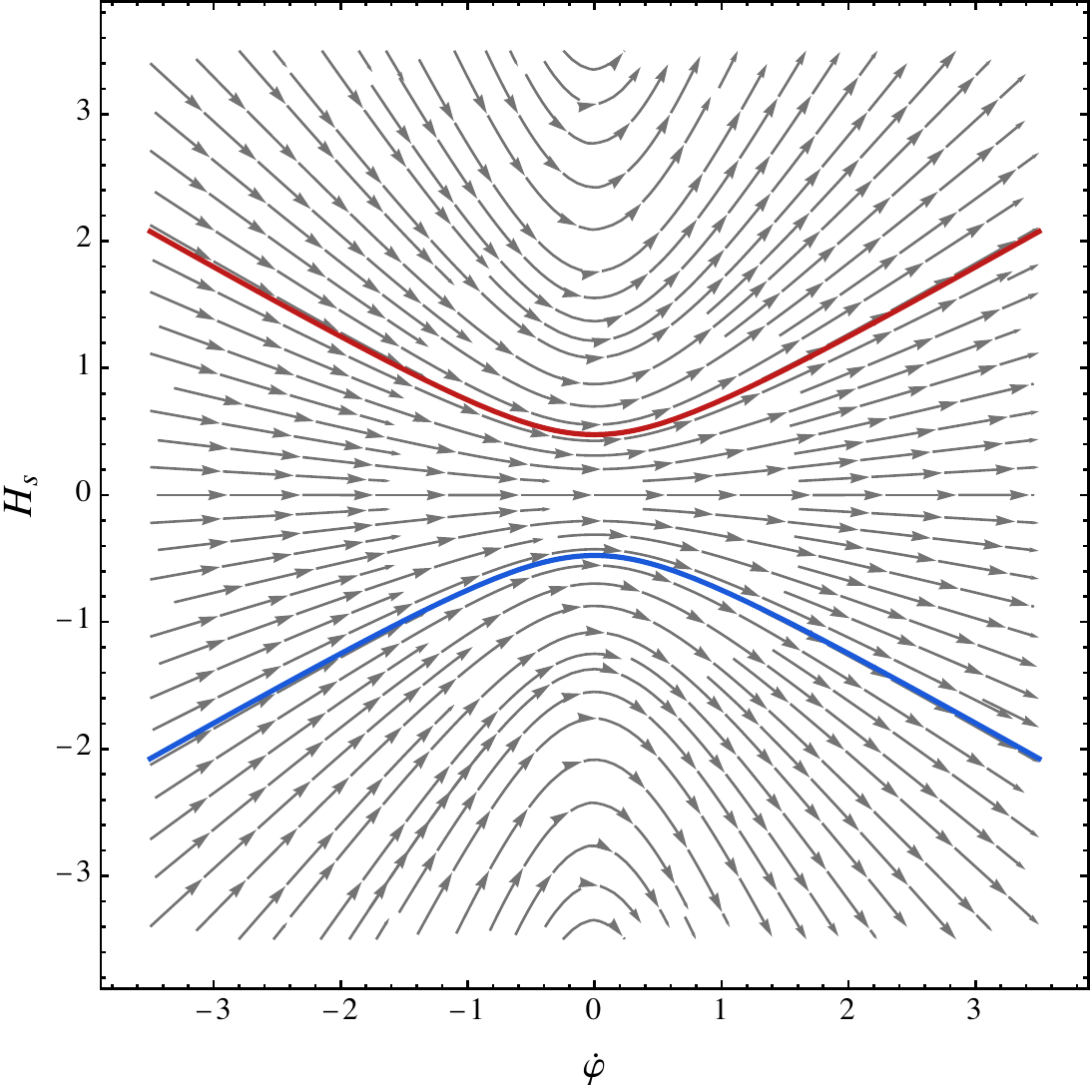}
    \caption{Phase portrait in the $(\dot\varphi, H_s)$ plane
    for $\mu^2 < 0$. The red and blue curves correspond to the Hamiltonian constraint; neither touches the $H_s = 0$
    axis.}
    \label{fig:tachyon-phase}
\end{figure}

The flow $\ddot\varphi = \dot\varphi^2 + |\mu^2|\chi^2$ has no fixed points. The energy density $\rho_\chi = \dot\chi^2 - |\mu^2|\chi^2$ is negative whenever $\dot\chi^2 < |\mu^2|\chi^2$, and the constraint becomes $3H_s^2 = \dot\varphi^2 - \dot\chi^2 + |\mu^2|\chi^2$. The constraint curves are hyperbolas with minimum Hubble magnitude $|H_s|_{\min} = \sqrt{|\mu^2|\chi^2/d} > 0$, so the $H_s = 0$ axis is never reached. Trajectories on the upper branch have $H_s > 0$ throughout while $\dot\varphi$ evolves monotonically from negative to positive: the universe transitions from Class~III to Class~I without crossing $H_s = 0$. This is a \emph{branch change}, not a bounce --- the Hubble rate never vanishes, but the dilaton velocity flips sign.

The mechanism of branch change driven by negative effective energy density — the ``Egg'' of Brustein and Veneziano~\cite{Brustein_1994} — applies here with the thermal scalar
providing an effective realization of the source
for the negative energy, with $\mu^2(\beta)$ fixed by string thermodynamics rather than postulated. Since $\rho_\chi + P_\chi  = 2 \dot\chi^2 \ge  0$ NEC is preserved throughout, consistent with~\cite{Kaloper:2007pw} that branch changes $(-) \to (+)$ require only negative energy density, not NEC violation.

To obtain an analytic expression for the branch-crossing timescale, we restrict to configurations with $\dot\chi = 0$ --- a simplifying choice that decouples the thermal scalar dynamics and reduces the dilaton equation to a single ODE. With $b \equiv |\mu|\chi$, the equation $\ddot\varphi = \dot\varphi^2 + b^2$ is solved by $ \dot\varphi(t) = b\tan\!\left[b(t-t_0)\right],\quad |t - t_0| < \frac{\pi}{2b}$ so the dilaton velocity sweeps from $-\infty$ to $+\infty$ in cosmic time $\Delta t = \frac{\pi}{|\mu|\chi}.$

The constraint gives $3H_s^2 = b^2\sec^2[b(t-t_0)]$, strictly positive throughout the interval; the trajectory is physical in the interior and diverges only at the endpoints. The timescale $\Delta t$ diverges as $|\mu| \to 0$: the branch crossing becomes slow as $T \to T_H^+$. We emphasize that $\Delta t$ is a statement about the gravi-dilaton response at a fixed thermal-scalar background; it is not a prediction for the dynamics of the ensemble itself.

We note that the tachyonic regime is where the Euclidean caveat discussed in Section~\ref{sec:thermalscalar} is most acute: the thermal scalar is an order parameter for the Hagedorn transition, and its tachyonic rolling in Lorentzian signature is precisely the behavior that Barbon and Rabinovici~\cite{BARB_N_2005} caution against interpreting as real-time dynamics. The branch-change mechanism identified here should therefore be understood as a feature of the effective equations rather than a controlled prediction. A deeper understanding of this regime likely requires going beyond the effective description used here.

\subsection{At the Hagedorn temperature ($\mu^2 = 0$)}
\label{sec:critical}

\begin{figure}[htbp]
    \centering
    \includegraphics[width=0.7\linewidth]{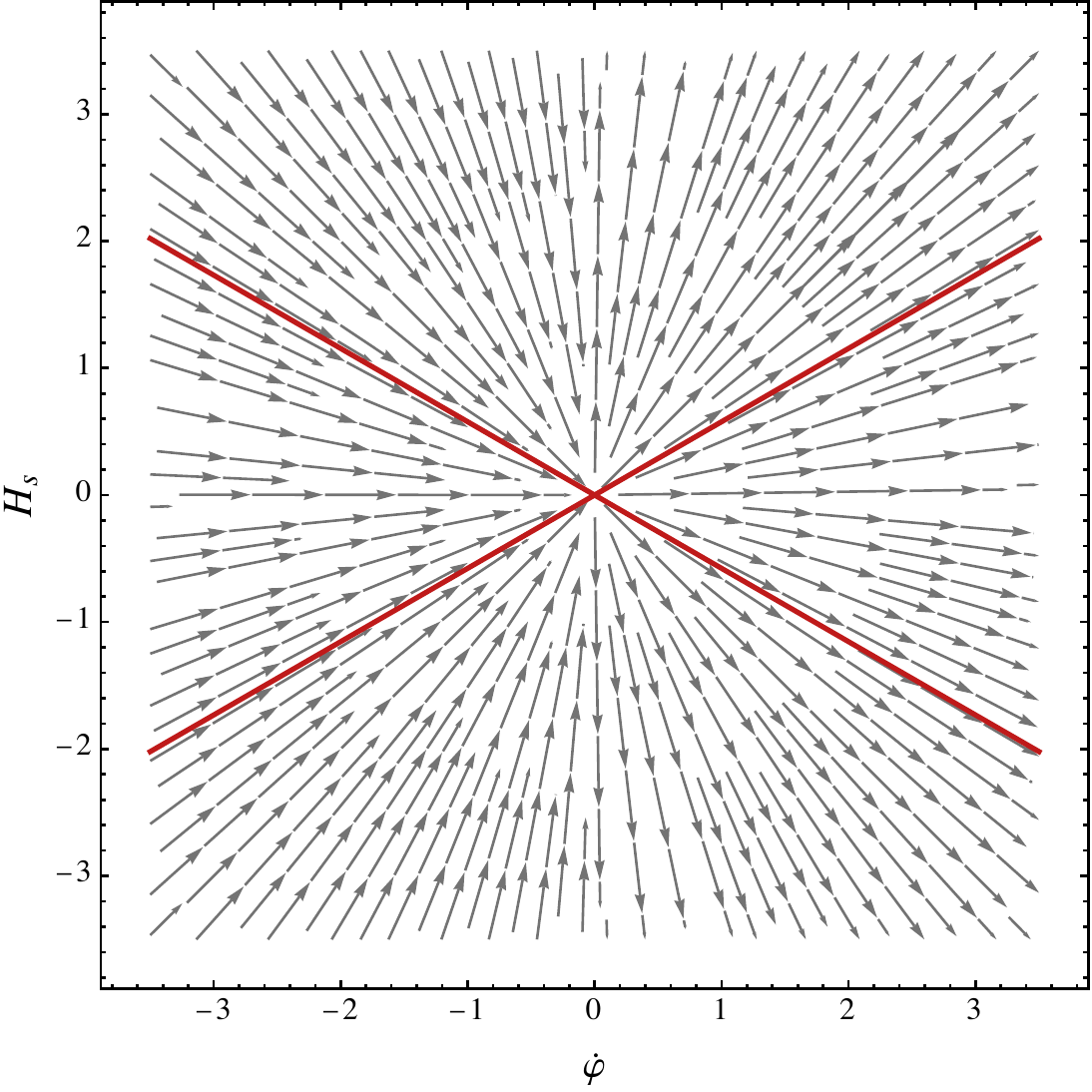}
    \caption{Phase portrait in the $(\dot\varphi, H_s)$  plane at $\mu^2 = 0 $. The two fixed points of the massive regime collapse to the origin. The red curves show the constraint $dH_s^2 = \dot\varphi^2 - \dot\chi^2$  for $\dot\chi=0$; all trajectories pass through the origin, which acts as a degenerate saddle.}
    \label{fig:placeholder}
\end{figure}

At $\mu^2 = 0$, the two fixed points of the massive regime at $(\pm\mu|\chi|, 0)$ merge at the origin, and the constraint reduces to $3H_s^2 = \dot\varphi^2 - \dot\chi^2$. Beyond this observation, there is little to say within the effective description used here.

The quadratic action becomes insufficient at $\mu^2 = 0$: the potential term vanishes, and the quartic coupling~\cite{Brustein_2021} — subleading in our truncation — becomes the leading contribution. A further caveat comes from the thermal ensemble interpretation itself. Atick and Witten~\cite{Atick:1988si} argued that the notion of temperature breaks down near the Hagedorn transition, as the latent heat is large enough that the thermal bath approximation fails. The limitations of using $\chi$ as a dynamical field, noted in Section~\ref{sec:thermalscalar}, become acute here.

We therefore do not make claims about the $\mu^2 = 0$ point itself. The results of Sections~\ref{sec:massive} and~\ref{sec:tachyonic} should be read as statements about the approach to $T_H$ from below and above, respectively.

\section{Momentum Modes}
\label{sec:momentummodes}

A natural question is whether an analogous field can be derived for the string momentum modes. Setting $w = 0$ and $k = \pm 1$ in the bosonic string mass spectrum yields a scalar with temperature-dependent mass
\begin{equation}
    \mu^2_\psi(\beta) = \frac{4\pi^2}{\beta^2} 
    - \frac{4}{\alpha'}\,,
    \label{eq:psimass}
\end{equation}
which is the T-dual of the thermal scalar mass. However, this mode is tachyonic at the Hagedorn temperature $T_H = 1/(4\pi\sqrt{\alpha'})$, with $\mu^2_\psi(\beta_H) = -15/4\alpha'$, due to the zero-temperature bosonic tachyon. It is therefore not a candidate for a light degree of freedom near the Hagedorn transition.

A different identification was made by Brustein and Zigdon~\cite{Brustein_2021}, who observed that the $k = \pm 4$ momentum mode becomes massless precisely at $T_H$, with mass
\begin{equation}
    m^2_{T_4} = 64\pi^2(T^2 - T_H^2)\,.
    \label{eq:T4mass}
\end{equation}
Since $\chi$ carries winding number and $T_4$ carries momentum, a quadratic mixing term $\chi^* T_4$ would violate winding-number conservation; the lowest interaction that preserves winding and momentum number separately is the quartic vertex $|\chi|^2|T_4|^2$. The combined action takes the schematic form~\cite{Brustein_2021}

\begin{equation}
    S_{\chi, T_4} \supset \int d^dx\,\sqrt{G}\, e^{-\phi}
    \left(|\partial\chi|^2 + \mu^2_\chi |\chi|^2 + g|\chi|^4
    + |\partial T_4|^2 + m^2_{T_4}|T_4|^2 + g|T_4|^4
    + 4g\,|\chi|^2|T_4|^2\right),
    \label{eq:BZaction}
\end{equation}
with $g = c_2 \kappa^2 T^2$~\cite{Brustein_2021}.

This construction is not included in the present analysis for three reasons. First, the winding and momentum symmetries forbid any quadratic cross-term, so at the quadratic level $\chi$ and $T_4$ decouple entirely: $T_4$ does not appear in any of the equations studied in Section~\ref{sec:thermalscalar}. Including its effects requires the full quartic action, where the thermal scalar self-coupling $|\chi|^4$ also becomes relevant. Second, the $k = \pm 4$ mode relies on the zero-temperature bosonic tachyon for its mass cancellation at $T_H$; in superstring theory, where there is no zero-temperature tachyon, there is no analogous momentum-mode partner that becomes massless at $T_H$~\cite{Brustein_2021}. Third, $T_4$ captures only the $k = \pm 4$ channel of the loop-production process; the full Brandenberger--Vafa mechanism involves winding strings annihilating into string loops carrying all momentum modes, and a complete treatment would require summing over them~\cite{Brandenberger:1988aj, Brandenberger_2002}.

The quadratic thermal scalar action therefore captures the dynamics of the approach to the Hagedorn transition but not the exit from it. Extending the analysis to the quartic action~\eqref{eq:BZaction} would allow the dominant winding annihilation channel to be studied dynamically in the gravi-dilaton background; we leave this for future work.

\section{Discussion and Outlook}
\label{sec:discussion}

In this paper, we consider the Hagedorn exit which is not a single obstruction but a combination of three distinct dynamical obstacles, each associated with a different regime of the thermal scalar effective theory. By coupling the thermal scalar to the string-frame gravi-dilaton system and analyzing the resulting phase space, we isolate the role played by each regime and identify where the limitations of the effective description arise.

The first obstruction appears below the Hagedorn temperature. In this regime, the thermal scalar admits static configurations in which its potential energy balances the shifted dilaton evolution, providing a field-theoretic realization of the lingering states discussed in String Gas Cosmology. However, these configurations are boundary states rather than attractors. When the thermal-scalar mass depends on the scale factor, winding-mode back-reaction opposes expansion and drives the cosmology toward contraction and increasing curvature. Thus, while winding modes can support stalled configurations, they do not naturally generate a sustained expanding phase. Instead, the dynamics are pushed toward a regime in which higher-curvature corrections become important and the low-energy effective description breaks down.

The second obstruction appears above the Hagedorn temperature. In the tachyonic regime, the thermal scalar generates a negative contribution to the effective energy density, allowing branch changes of the Brustein–Veneziano type while preserving the null energy condition. Within the effective theory, this provides a concrete realization of the conditions required for branch changing. However, the resulting transitions proceed from the ((-)) branch to the ((+)) branch, whereas a successful graceful exit requires the reverse transition. The tachyonic regime therefore demonstrates that branch changes are possible, but it does not produce the particular branch change required to connect the Hagedorn phase to the standard cosmological evolution observed today.
The third obstruction occurs precisely at the Hagedorn temperature itself. As the thermal scalar becomes massless, the fixed points of the massive regime collapse and the quadratic approximation loses predictive power. At the same time, the quartic interactions neglected in the present analysis become equally important, and the thermal ensemble interpretation underlying the effective description begins to fail. The transition itself therefore sits at the boundary of validity of the quadratic thermal-scalar theory. While the approach to the transition can be studied within the effective description, the transition and its completion cannot.

Viewed together, these three obstructions provide a unified picture of the Hagedorn exit problem. Below ($T_H$), winding modes resist expansion and drive the system toward strong curvature. Above ($T_H$), branch changes become possible but occur in the wrong direction. At ($T_H$), the effective theory itself breaks down. The exit problem therefore lies precisely at the intersection of all three effects. This explains how it has proven difficult to realize a successful Hagedorn exit within controlled effective descriptions -- every regime supplies part of the required physics, but none contains all of it.

From this perspective, the thermal scalar does not solve the Hagedorn exit problem, but it clarifies and gives a new approach to understand the structure of the problem. The quadratic thermal-scalar effective theory provides a description of the approach to the Hagedorn transition from both sides and reveals which ingredients are missing from the exit mechanism. Any successful resolution must incorporate winding-number violation, interactions between winding and momentum sectors, and dynamics beyond the quadratic truncation.

Several directions for future work naturally follow. The most immediate is the quartic effective theory of Brustein and Zigdon, which introduces both thermal-scalar self-interactions and couplings to the momentum sector. These interactions provide the leading effective field description of winding annihilation near the Hagedorn temperature and offer a framework in which the exit mechanism may be studied dynamically. A second direction is deriving the scale-factor dependence of the thermal-scalar mass directly from string thermodynamics rather than through phenomenological ansätze. Finally, it would be interesting to investigate whether analogous dynamics arise in superstring constructions, where the absence of the bosonic tachyon modifies the structure of the momentum sector and may alter the physics of the transition itself. We leave all of this to future work.

\section*{Acknowledgements}
We thank Nemanja Kaloper for past collaboration and discussion, and Alex Maloney, Keith Dienes, and Damien Easson for useful conversations. 
This research was supported in part by DOE grant DE-FG02-85ER40237 and the Simons Center. AP acknowledges support from the Natural Sciences and Engineering Research Council (NSERC) of
Canada.

\bibliography{paper.bib}
\end{document}